\documentclass[journal]{IEEEtran}
\usepackage{cite}
\usepackage{amsmath,amssymb,amsfonts}
\usepackage{algorithmic}
\usepackage{graphicx}
\usepackage{textcomp}
\usepackage[export]{adjustbox}
\usepackage{xcolor}
\usepackage{gensymb}
\usepackage{listings}
\usepackage{url}

\usepackage[margins]{trackchanges}
\ifCLASSINFOpdf
\else
\fi
\begin{document}
%
\title{Analysis of Trade-offs in RF Photonic Links based on Multi-Bias Tuning of Silicon Photonic Ring-Assisted Mach Zehnder Modulators}
%
%
%

\author{Md~Jubayer Shawon,~\IEEEmembership{Student Member,~IEEE,}
        and~Vishal~Saxena,~\IEEEmembership{Senior Member,~IEEE}
\thanks{The authors are with the Department of Electrical and Computer Engineering, University of Delaware, Newark,
DE, 19716 USA e-mail: shawon@udel.edu.}
}

\maketitle
\thispagestyle{empty}
\pagestyle{empty}

\begin{abstract}
Recent progress in silicon-based photonic integrated circuits (PICs) have opened new avenues for analog circuit designers to explore hybrid integration of photonics with CMOS ICs. Traditionally, optoelectronic systems are designed using discrete optics and electronics. Silicon photonic (SiP) platforms provide the opportunity to realize these systems in a compact chip-scale form factor and alleviate long-standing challenges in optoelectronics. In this work, we analyze multi-bias tuning in Ring-Assisted Mach Zehnder Modulator (RAMZM) and resulting trade-offs in analog RF photonic links realized using RAMZMs. Multi-bias tuning in the rings and the Mach-Zehnder arms allow informed trade-offs between link noise figure and linearity. We derive performance metrics including gain, noise figure, and linearity metrics associated with tuning of multiple bias settings in RAMZM based links and present resulting design optimization. Compared to MZM, an improvement of 18 dB/Hz$^{\frac{2}{3}}$ in SFDR is noted when RAMZM is linearized.  We also propose a biasing scheme for RAMZM that provides 6x improvement in slope efficiency, or equivalently, 15.56dB in power Gain over MZMs (single drive) while still providing similar SFDR performance ($\sim$ 109 dB/Hz$^{\frac{2}{3}}$) as MZMs. Moreover, a method to improve gain in photodiode saturation limited links is presented and studied.
\end{abstract}

\begin{IEEEkeywords}
Analog optical link, Silicon Photonics, Photonic Integrated Circuit (PIC), Ring-Assisted Mach Zehnder Modulator (RAMZM), RF Photonics, RF-to-optical modulator.
\end{IEEEkeywords}

%
\IEEEpeerreviewmaketitle

\section{Introduction}
%
%
%
%
\IEEEPARstart{T}{he} emergence of silicon-based photonic integrated circuits (PICs) and the ability to fabricate them on foundry multi-project wafer (MPW) platforms is now allowing analog circuit designers to integrate entire optoelectronic system in the same chip-scale package. Radio-frequency (RF) photonics is an established field where RF signals are processed and then transmitted over optical fibers using discrete lasers, modulators (photo)detectors and associated electronics. In addition to improved size, weight, power, and cost (SWaPc) metrics, large-scale PICs can be designed to alleviate longstanding limitations of the discrete optoelectronic systems. 

\begin{figure}[!t]
\centering
\label{fig:analog_link}
\includegraphics[width=\columnwidth]{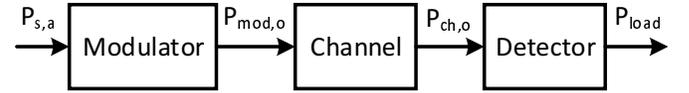}
\caption{An analog optical link using RAMZI modulator. Here, $P_I$ and $P_{M,o}$ are the input and output optical power at the modulator, and $P_{o,d}$ is the optical input power to the detector. $P_{s,a}$ and $P_L$ are the available input power and output powers of the link.}
\label{fig:RAMZI_link} 
\end{figure}

As shown in Fig. \ref{fig:analog_link}, an analog RF photonic optical link employs either a direct modulated laser (DML) or a Mach-Zehnder Modulator (MZM) which  modulates a continuous-wave (CW) laser source (typically at the 1550nm telecommunication wavelength)  using an RF signal. The RF signal in the optical domain undergoes optical signal processing (e.g. filtering, phased-array processing, etc.) and transmitted over optical fiber(s). The optical-domain RF signal is finally converted back into RF domain using a (photo)detector, which is optionally followed by a low-noise transimpedance amplifier (LNTIA). 
In Fig. \ref{fig:analog_link}, $P_{s,a}$ is the input RF source power, $P_{M,o}$ is the optical power at the modulator output, $P_{o,d}$ is the optical power at the detector input, $P_{L}$ is the RF power at the load. Since the detected RF power depends upon the square of optical power, the optical gain (loss) is squared in the link gain ($G$), determined as
\begin{equation} \label{eq:LinkGain}
	G = G_{M} \cdot G_{ch}^2 \cdot G_{det}
\end{equation}
where $G_{M} \triangleq \frac{P_{M,o}^2}{P_{s,a}}$, $G_{ch} \triangleq \frac{P_{o,d}}{P_{M,o}}$, $G_{det} \triangleq \frac{P_L}{P_{o,d}^2}$ are the individual available power gains (or losses) for the modulator, optical channel, and the detector respectively. 

In prior art, the whole link is modeled end-to-end and the overall small-signal gain, noise figure (NF), and distortion metrics are determined \cite{cox2006analog,ackerman2003effective}. Though MZM-based links have been here for the past several decades, their analog performance is limited by their inherent raised-cosine optical non-linearity.  

Recently, ring-assisted Mach-Zehnder modulator (RAMZM) has been investigated with significantly improved linearity performance\cite{mwscas18a,cardenas2013linearized}. However, the performance of RAMZM in an analog photonic link has not been studied in detail. The authors provided a preliminary analysis in \cite{shawon2020analysis} and this article expands on this analysis while incorporating novel design trade-offs based on the tuning of multiple optical biasing in the RAMZM. 

In this article, we present analysis for analog RF photonic links using RAMZM for high linearity. We derive the small-signal gain, noise figure, and linearity metrics associated with the RAMZM based link and present design trade-offs. The rest of this article is organized as follows: Section II describes the RAMZM and derives small-signal gain expression. In section III, the noise figure of the entire optical link is derived and simulated. Section IV summarizes link distortion analysis followed by conclusion.

\section{Ring-Assisted Mach-Zehnder Modulator}

\subsection{MZM and Microring Modulators}

A linear RF-to-optical modulation is required before optical-domain signal processing is performed. Discrete lithium niobate (LiNbO$_3$) MZMs have been widely used in analog photonic links and are limited to $<95$ dB/Hz$^{2/3}$ spur-free dynamic range (SFDR) at 1GHz. Linearization of MZMs has been a long-studied topic in RF photonics \cite{cox2006analog,urick2015fundamentals}. Active electronic linearization using a CMOS pre-distortion circuit has been investigated using third or fifth-order active predistortion \cite{sadhwani2003adaptive, okyere2017fifth}. The active linearization techniques exhibit lower bandwidths ($<1$GHz), and incur noise penalty and supply voltage limitation. A third-order intermodulation (IM$_3$) cancellation technique was recently demonstrated with $<120$ dB/Hz$^{2/3}$ SFDR at 9GHz \cite{hosseinzadeh2020distributed}. However, this technique is narrowband around the 9GHz center frequency. 

Silicon photonics based analog MZMs have been designed where DC Kerr non-linearity was used to compensate for the pn-junction non-linearity to achieve a broadband SFDR linearity of $\sim 106$ dB/Hz$^{2/3}$  over a 13GHz bandwidth \cite{bottenfield2019silicon,timurdogan2017electric}. However, such analog modulators require a long traveling-wave electrode design ($<2mm$) which incurs higher noise figure due to electrode losses \cite{timurdogan2019apsuny}.

Microring modulators (MRM) allow modulation with a very small form-factor ($\sim 5\mu m$ radius) and lumped capacitive interface, but typically suffer both even as well as odd-order distortion due to the Lorentzian optical transmission of the microring. Consequently, SiP MRMs exhibit an SFDR of $63$ dB/Hz$^{2/3}$ with single-ended drive \cite{ayazi2012linearity}.  Linearization of MRMs has been attempted but the resulting performance is limited below $98$ dB/Hz$^{2/3}$ up to 13 GHz \cite{jain2019high}.

\begin{figure}[!tbh]
\centering
\includegraphics[width=\columnwidth]{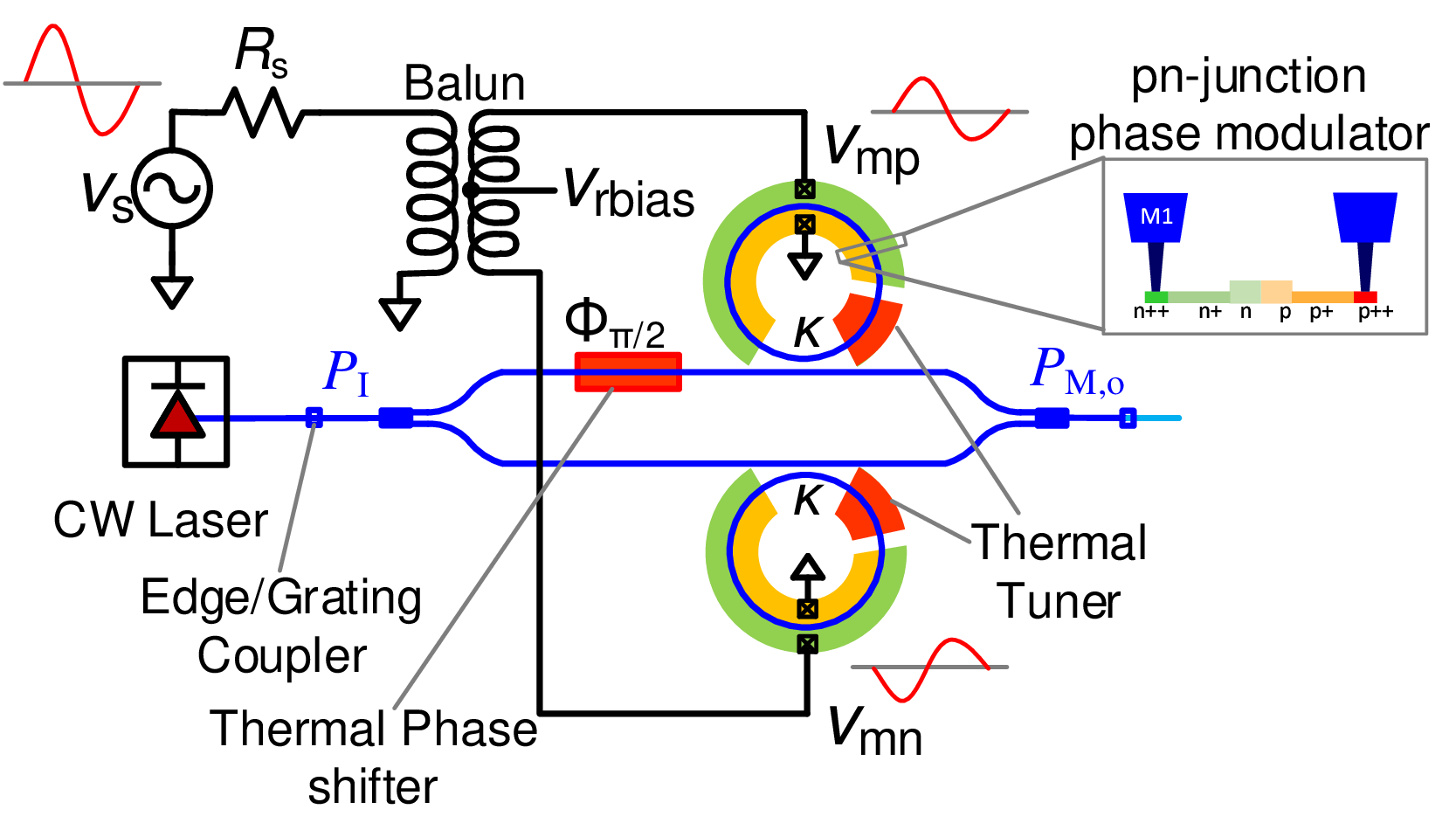}
\caption{Conceptual schematic of a silicon photonic Ring Assisted Mach Zehnder Modulator (RAMZM), interfaced with an RF source and a CW laser source. }
\label{fig:ramzi}
\end{figure}

\subsection{RAMZM Modulators}

In the last decade, ring-assisted linearized MZ (RAMZI) modulators have been explored where a differential configuration, similar to MZM, is employed but the two (upper and lower) phase shifters are replaced by ring modulators as shown in Fig. \ref{fig:ramzi}. By exploiting the Lorentzian transmission characteristics, upper and lower rings can be optically biased such that their voltage-to-phase responses are expansive. This expansive response can be enhanced by a suitable value of coupling coefficient ($\kappa$) and then used to compensate for the compressive phase-to-intensity non-linearity of a Mach-Zehnder interferometer \cite{gutierrez2012ring,cardenas2013linearized,morton2013morton,mwscas18a}. The MZ push-pull configuration suppresses and even-order non-linearity. 

\begin{figure}[!tbh]
\centering
\includegraphics[width=\columnwidth]{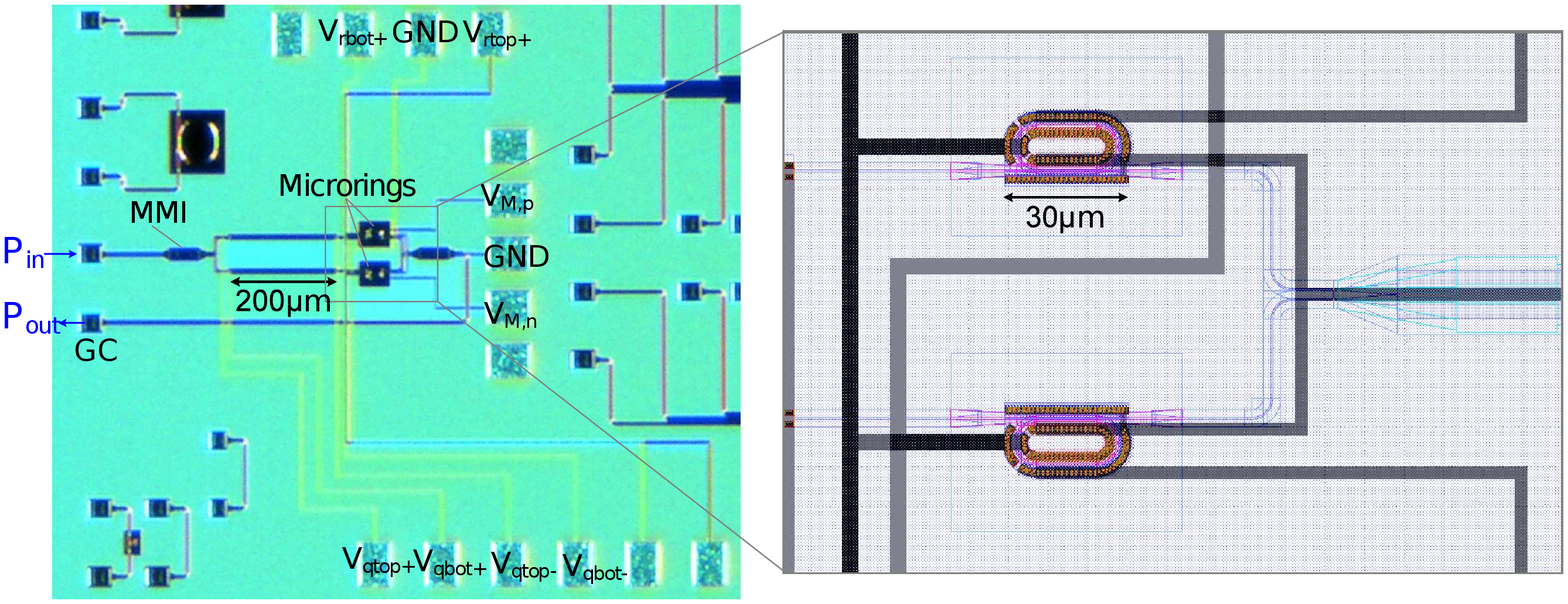}
\caption{Chip micrograph of a RAMZM fabricated in IMEC's ISIPP50G SiP process.}
\label{fig:ramzi_chip}
\end{figure}

Fig. \ref{fig:ramzi} depicts the schematic of a SiP RAMZM and its interface with the RF input source using a differential balun. Fig. \ref{fig:ramzi_chip} shows the chip micrograph of an RAMZM fabricated by the authors using IMEC's ISIPP50G SiP process \cite{imec50G}. As shown in Fig. \ref{fig:ramzi}, the output of the CW laser is equally split into two arms of the MZM. RF signals are applied to the ring modulators in a differential push-pull configuration around a DC bias ($V_{M}$). These differential RF signals are translated into different phase changes in each arms of the MZI. When the two optical waves combine at the output of the MZM, optical modulation is achieved via constructive or destructive interference of light \cite{chrostowski2015silicon}.


\subsection{RAMZM Transmission Characteristics}
We now derive the transmission characteristics of the RAMZM. The optical wave electric field at the output of a RAMZM can be expressed as \cite{yue2013mmi} 

\begin{equation} \label{eq:ein}
\begin{aligned}
E_{out} &= \frac{E_{in}}{2} \Big( \left | a_{r1}(\theta) \right |e^{-j.(knL_1+ \phi_{bias} +\angle a_{r1}(\theta))} + \cdots \\ 
&+ \left | a_{r2}(\theta) \right |e^{-j.(knL_2+ \angle a_{r2}(\theta))} \Big)
\end{aligned}
\end{equation}

\begin{equation} \label{eq:artht}
\begin{aligned}
a_{ri}(\theta) = \frac{\tau - \alpha e^{-j\theta}}{1 - \tau \alpha e^{-j\theta}}
\end{aligned}
\end{equation}

where the variables are defined below. Here, index term $i=1$ refers to upper while $i=2$ refers to the lower arm of RAMZM. \\ 
$E_{in}$: input electric field\\
$L_i$: length of the RAMZM arms\\
$\left | a_{ri}(\theta)\right |$: microring magnitude response\\
$\angle a_{ri}(\theta)$: microring phase response\\
$\phi_{bias}$: phase difference between the upper and lower arms introduced by thermal phase shifter\\ 
$\tau=\sqrt{1-\kappa^2}$: transmission coefficient of the microring couplers \\
$\alpha$: microring loss factor\\
$\theta=\theta(V)$: microring roundtrip phase shift, which depends on the applied voltage in the phase-modulators. \\

If we neglect losses in the short MZI arms, the transmission function, $T(\theta) = \frac{I_{out}}{I_{in}}= \Big|\frac{E_{out}}{E_{in}}\Big|^2$, can be written as


\begin{align} \label{eq:tf}
T(\theta) 
&= \frac{1}{4}\left|e^{-j.(knL_1+\phi_{bias} + \angle a_{r1}(\theta))} + e^{-j.(knL_2+ \angle a_{r2}(\theta))} \right|^2
\end{align}


When the optical path of the RAMZM arms have equal length ($L_2 = L_1$), the transfer function can be simplified as  

\begin{align}
T(\theta) &= \frac{1}{2}\left [ 1+\cos \Big(\phi_{bias}+\angle a_{r1}(\theta)-\angle a_{r2}(\theta) \Big) \right ] \label{eq:tf2} 
\end{align}

\begin{equation} \label{eq:aris}
\begin{aligned}
\phi_i = \angle a_{ri}(\theta) = \tan^{-1}\left ( \frac{\alpha (1-\tau^2)\sin\theta}{\tau(1+\alpha^2)-\alpha (1+\tau^2)\cos\theta} \right )
\end{aligned}
\end{equation}

As mentioned earlier, $\theta$ is the round-trip phase delay in the microrings. By applying differential signal ($\theta = \theta_{DC} \pm \theta_{mod}$) in the microrings, one can achieve electro-optic modulation. Here, $\theta_{DC}$ is the microring optical DC bias point, whereas $\theta_{mod}$ is the modulating RF input signal. Now, if we assume low-loss rings, i.e. $\alpha \approx 1$, and apply differential modulating signal, trigonometric manipulation of $T(\theta)$ and expanding its series around $\theta_{mod}$ gives us Eq. \ref{eq:Pmo1} to Eq. \ref{eq:Pmo1_last} \cite{cox2006analog,marpaung2009high}.

\newcounter{mytempeqncnt}

\begin{figure*}[!t]
\setcounter{mytempeqncnt}{\value{equation}}
\setcounter{equation}{11}

\begin{equation} \label{eq:Pmo1}
P_{M,o}(\theta_{mod}) = P_I T(\theta)= \frac{P_I \gamma_0}{2} - \gamma_1 P_I \theta_{mod} -\gamma_2 P_I \theta_{mod}^2+ \gamma_3 P_I \theta_{mod}^3 + \gamma_4 P_I \theta_{mod}^4 + \mathcal{O}(\theta_{mod}^5)
\end{equation}

\begin{align}
\gamma_0 &= 1+\cos(\phi_{bias}) \\
\gamma_1 &= \frac{(\tau^2-1)\cdot \sin(\phi_{bias})}{\tau^2-2\tau \cos(\theta_{DC})+1} \\
\gamma_2 &= \frac{(\tau^2-1)^2\cdot \cos(\phi_{bias})}{(\tau^2-2\tau \cos(\theta_{DC})+1)^2} \\
\gamma_3 &= \frac{(\tau^2-1)(2\cos(\theta_{DC})^2 \tau^2 + \cos(\theta_{DC}) \tau^3 + 2\tau^4 + \cos(\theta_{DC}) \tau -8\tau^2 +2)\cdot \sin(\phi_{bias})}{3(\tau^2-2\tau \cos(\theta_{DC})+1)^3} \\
\gamma_4 &= \frac{(\tau^2-1)^2 (4\cos(\theta_{DC})^2 \tau^2 + 2\cos(\theta_{DC}) \tau^3 + \tau^4 + 2\cos(\theta_{DC}) \tau -10\tau^2 +1)\cdot \cos(\phi_{bias})}{(\tau^2-2\tau \cos(\theta_{DC})+1)^4}  \label{eq:Pmo1_last}
\end{align}

\setcounter{equation}{\value{mytempeqncnt}+6}
\hrulefill
\vspace*{4pt}
\end{figure*}




In these equations, $P_I$ is the modulator optical input power. Upon observing Eq. \ref{eq:Pmo1}, we can easily deduce that even-order distortions can entirely be eliminated simply by setting $\phi_{bias} = \frac{\pi}{2}$. This is achieved by tuning thermal heaters on the RAMZM arms \cite{chrostowski2015silicon}. On the other hand, the third-order nonlinearity can be cancelled by setting the rings at the anti-resonance point (i.e. optical bias $\theta_{DC} = \pi$) \cite{cardenas2013linearized}. This is achieved by tuning the thermal phase shifters of the microrings while the driver voltage is set to a common-mode voltage, $V_M$ DC. The differential modulating signals, $v_{m}$, are applied around this common-mode voltage, so that the individual ring drive voltages are $V_M \pm v_{m}$. Therefore, the round-trip phase shift at the upper and lower ring will be $\pi+\theta_{mod}(v_m)$ and $\pi-\theta_{mod}(v_m)$, respectively.

\begin{figure}[!h]
\centering
\includegraphics[width=0.9\columnwidth]{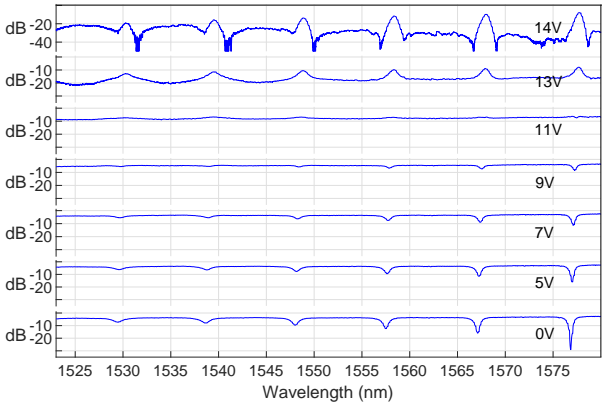}
\caption{Spectral response of RAMZM (fabricated in IMEC’s ISIPP50G SiP process) when arm bias ($\phi_{bias}$) is tuned by applying 0-14V DC on a $n$-doped resistive heater placed around the arm.}
\label{fig:quadbias}
\end{figure}

\subsection{RAMZM Optical Biasing}

As mentioned before, biasing schemes are implemented on a SiP RAMZM by means of thermal phase shifters. The SiP RAMZM presented in this work utilizes $n$-doped resistive heaters \cite{jayatilleka2015wavelength} placed around the arms and rings to tune $\phi_{bias}$ (arm bias) and $\theta_{DC}$ (ring bias), respectively. The change in spectral response of the RAMZM with various arm bias voltages (0-14V DC) is shown in Fig. \ref{fig:quadbias}. Here, a laser source was coupled into the on-chip RAMZM by means of a fiber array and grating coupler \cite{chrostowski2015silicon}. The RAMZM output was taken off the chip via another grating coupler and fiber array. The fiber array was terminated at a butt-coupled photodetector.

\begin{figure}[!h]
\centering
\includegraphics[width=0.9\columnwidth]{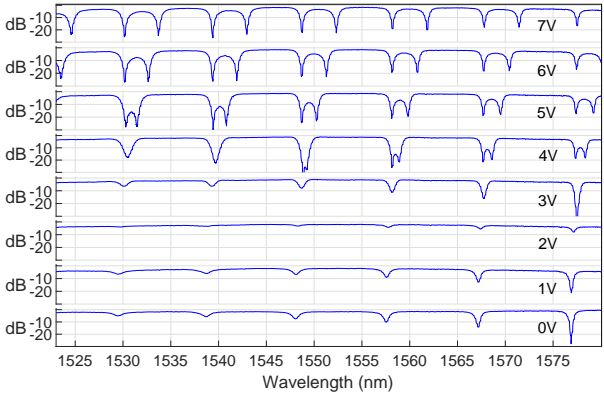}
\caption{Spectral response of RAMZM (fabricated in IMEC’s ISIPP50G SiP process) when lower microring bias ($\theta_{DC}$) is tuned by applying 0-7V on a $n$-doped resistive heater placed around the ring.}
\label{fig:lowerringbias}
\end{figure}

\begin{figure}[!h]
\centering
\includegraphics[width=0.9\columnwidth]{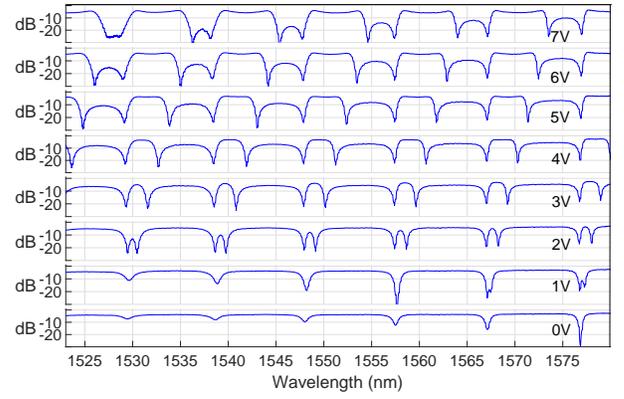}
\caption{Spectral response of RAMZM (fabricated in IMEC’s ISIPP50G SiP process) when upper microring bias ($\theta_{DC}$) is tuned by applying 0-7V on a $n$-doped resistive heater placed around the ring.}
\label{fig:upperringbias}
\end{figure}

Similarly, by applying 0-7V DC on the ring heaters, the resonances of the lower (Fig. \ref{fig:lowerringbias}) and upper (Fig. \ref{fig:upperringbias}) ring were tuned. Here, it can be seen that as the tuning voltage is increased, multiple resonance peaks start appearing simply due to the resonance mismatch between the upper and lower rings. To linearize RAMZM, both microrings need to have their resonances matched. This can easily be achieved by `detuning' one or both of the rings \cite{zhang2016ultralinear}. It is important to note that the spectral response shown in Fig. \ref{fig:quadbias} - Fig. \ref{fig:upperringbias} has been deembedded and normalized for the optical response of grating couplers.

\begin{figure}[!h]
\centering
\includegraphics[width=1\columnwidth]{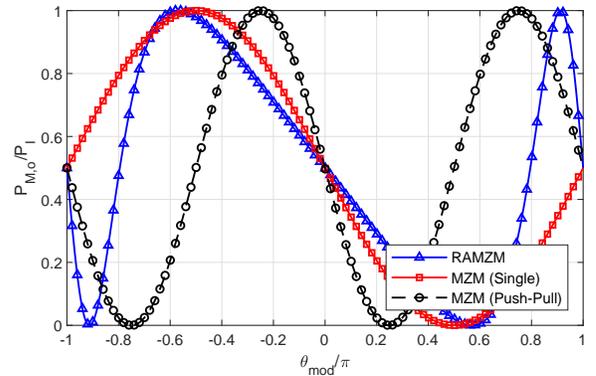}
\caption{RAMZM, MZM (single drive) and MZM (push-pull drive) Optical Transfer Function as a function of RF voltage induced relative phase shift, $\theta_{mod}$. Here, the RAMZM is biased at $\{\phi_{bias}, \theta_{DC}, \tau \}=\{\frac{\pi}{2},\pi, \frac{1}{2}\}$ and both MZMs are quadrature biased.}
\label{fig:xfr}
\end{figure}

The aforementioned bias conditions give us the Taylor series coefficients $\gamma_1 = \frac{1-\tau}{1+\tau}$ and $\gamma_3 = \frac{2\tau^3 - 7\tau^2 + 7\tau - 2}{3(1+\tau)(\tau^2+2\tau+1)}$. Interestingly, the third-order nonlinearity can be set to zero by enforcing $\gamma_3 = 0$, which is achieved by setting $\tau=\frac{1}{2}$. This ensures that third-order distortion is suppressed, leaving the modulator with fifth or higher odd-order distortions, which are very miniscule in practical application scenarios. The optical power transfer function of a linearized $(\{\phi_{bias}, \theta_{DC}, \tau\} = \{\pi/2, \pi, 0.5\})$ RAMZM along with the transfer functions of MZMs (both single and push-pull drive \cite{zhu2015design}) are plotted in Fig. \ref{fig:xfr}. As can be seen, the RAMZM is significantly linear compared to the MZMs when biased at this regime.

\begin{figure}[!h]
\centering
\includegraphics[width=\columnwidth]{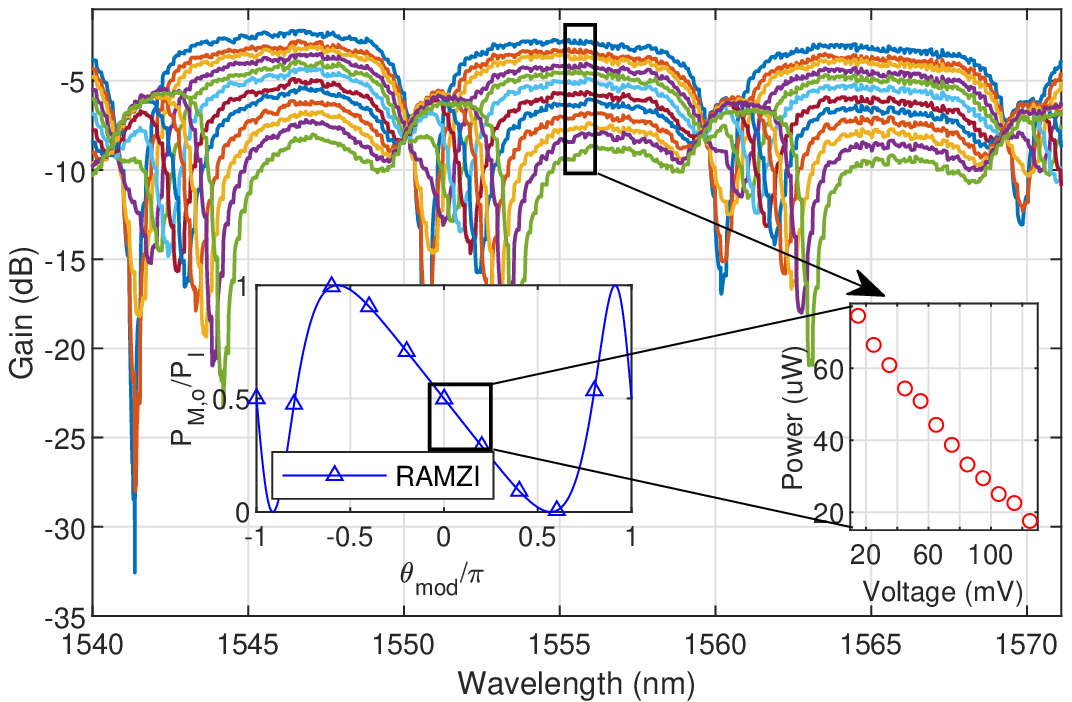}
\caption{RAMZM spectral response when both upper and lower ring p-n phase shifters are biased at $1.125V$ and applying modulation signal, $\theta_{mod}(v_m)$, on both microring p-n phase shifters in a differential manner. The inset plot on the right shows the modulation of photodiode output power based on $\theta_{mod}(v_m)$ at 1555.55nm. The left inset presents ideal RAMZM transfer function as a function of $\theta_{mod}(v_m)$.} 
\label{fig:rfmod}
\end{figure}

On chip, modulation signal is applied on the high speed p-n phase shifters placed around the microring waveguide as shown in Fig. \ref{fig:ramzi}. The input RF signal modulates the carriers present within the microring waveguide, thus modulating the phase of light. This phenomena is known as plasma dispersion effect \cite{chrostowski2015silicon}. When the antiresonance point ($\theta_{DC} = \pi$) is determined at a particular wavelength (1555.55nm in our case), the input signal will modulate the output optical power of RAMZM (as shown in the inset of Fig. \ref{fig:rfmod}). In this work, the p-n phase shifters on microrings are biased at 1.125V and differential modulating signal was applied on the rings. The corresponding detected photodiode power exhibits linear response.

\subsection{RAMZM Gain and Slope Efficiency}

Now, assuming that a plasma dispersion effect based (depletion or forward-bias mode) phase modulator with linear response is realized (combined effect of plasma dispersion and DC Kerr effect \cite{timurdogan2017electric}), we express the phase modulation as $ \theta_{mod} = \frac{\pi v_{m}}{V_{\pi}}$. Here, $V_{\pi}$ is defined as the voltage required to achieve $\pi$ phase shift in the given length of the phase modulator. Now, neglecting higher-order modulation terms, we obtain the RAMZM small signal output optical power as \cite{cox2006analog}-

\begin{equation} \label{eq:Pmo2}
P_{M,o} = \frac{\pi \gamma_1 P_I v_m}{V_{\pi} L}
\end{equation}

Here, the term `L' is used to include the effects of optical loss. If lossy impedance matching \cite{cox2006analog} at the input side (as shown in Fig. \ref{fig:Link_noise_model}) is implemented, the differential voltage across the modulator is related to the available power by

\begin{equation} \label{eq:Psa_vm}
    v_m = \frac{v_s}{2} = \sqrt{4P_{s,a}R_s}
\end{equation}

Consequently, the small-signal available power gain is given by \cite{cox2006analog} -

\begin{align} \label{eq:G_m_small_signal}
    G_M &= \frac{P_{M,o}^2}{P_{s,a}} 
        = \frac{s_{rmz}^2}{R_s}= \Big[\frac{\pi \gamma_1 P_I R_s}{V_{\pi} L} \Big ]^2 \frac{1}{R_s} 
\end{align}

where $s_{rmz} \triangleq \frac{P_{M,o}}{v_m} = \frac{\pi \gamma_1 P_I R_s}{V_{\pi} L} = \frac{\pi P_I R_s}{3V_{\pi} L}$ (when $\tau = \frac{1}{2}, \gamma_1 = \frac{1}{3}$) is the RAMZM slope efficiency. The effective RAMZM gain can be increased by increasing the input optical power $P_I$ (from a CW laser), reducing the insertion loss $L$, or decreasing modulator's $V_{\pi}$. To put that in perspective, MZM operating in singe and push-pull  drive has the slope efficiencies of $s_{mz} = \frac{\pi P_I R_s}{2V_{\pi} L}$ and $s_{mz} = \frac{\pi P_I R_s}{V_{\pi} L}$, respectively.

\subsection{Lumped Capacitive Drive}

Since, in most of the cases, RAMZM is treated as a lumped capacitive element from the electrical point of view, the voltage across the capacitance when driven using an external transmission line (of characteristic impedance $Z_0$) can be expressed as

\begin{equation} \label{eq:Gamma1}
    v_m = v_{+} + v_{-} = (1+|\Gamma_m|)v_s
\end{equation}
where 
\begin{equation} \label{eq:Gamma2}
    \Gamma_m = \frac{Z_m - Z_0}{Z_m + Z_0} 
             = \frac{r_m - Z_0 + \frac{1}{j\omega C_m}}{r_m +  Z_0 + \frac{1}{j\omega C_m}}
\end{equation}

is the frequency-dependent reflection coefficient. At low frequencies, $|\Gamma_m|\approx1$ and thus $v_m = 2v_{+} = 2(v_s/2) =v_s$. On the contrary, for a transmission-line drive MZM with impedance matching, we have $v_m = v_s/2$ \cite{cox2006analog}. Therefore, for a lumped capacitive drive the small-signal power gain is

\begin{align} \label{eq:G_m_small_signal}
    G_M &= \Big[\frac{2\pi \gamma_1 P_I R_s}{V_{\pi} L} \Big ]^2 \frac{1}{R_s} 
\end{align}

where $s_{rmz} \triangleq \frac{P_{M,o}}{v_m} = \frac{2\pi P_I R_s}{3V_{\pi} L}$.

Although lumped capacitive treatment of RAMZM yields better slope efficiency (2$\times$), we will use $G_M$ and $s_{rmz}$ derived in the previous subsection throughout the rest of this work for a fair comparison between RAMZM and MZM.


\subsection{Improving RAMZM Gain}

It's evident from the previous analysis and Fig. \ref{fig:xfr} that the linear response of RAMZM comes at the cost of the small-signal power gain. While RAMZM provides excellent distortion cancellation upto 5th order, its slope efficiency is only 67\% and 33\% of that of an MZM driven in single and push-pull manner, respectively. To improve the slope efficiency of the RAMZM, the derivative of the transfer function with respect to $\theta$ can be evaluated to find the bias point where slope efficiency is the maximum for a given $\tau$. Here, we propose a biasing scheme where $\theta_{DC} = 0$, instead of $\pi$ while keeping $\phi_{bias} = \frac{\pi}{2}$. The transfer function presented in Eq. \ref{eq:tf} can be simplified and expanded as

\begin{equation} \label{eq:Pmo1mod}
P_{M,o}(\theta_{mod}) = P_I T(\theta)= \frac{P_I}{2} - \gamma_1 P_I \theta_{mod} + \gamma_3 P_I \theta_{mod}^3 + \mathcal{O}(\theta_{mod}^5)
\end{equation}

\begin{figure}
\centering
\includegraphics[width=\columnwidth]{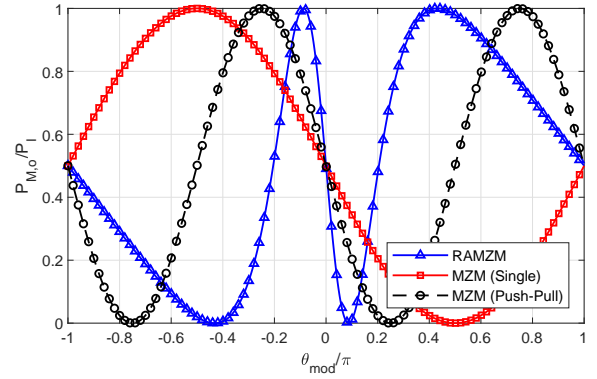}
\caption{RAMZM, MZM (single drive) and MZM (push-pull drive) Optical Transfer Function as a function of RF voltage induced relative phase shift, $\theta_{mod}$. Here, the RAMZM is biased at $\{\phi_{bias}, \theta_{DC}, \tau \}=\{\frac{\pi}{2},0,\frac{1}{2}\}$ and both MZMs are quadrature biased.}
\label{fig:highgain}
\end{figure}

Here, $\gamma_1 = \frac{1+\tau}{1-\tau}$ and $\gamma_3 = \frac{2\tau^3 + 7\tau^2 + 7\tau + 2}{3(\tau-1)(\tau^2-2\tau+1)}$. To optimize the gain and noise performance of the modulator, $\gamma_1$ needs to be maximized (i.e. $\tau\to 1$) whereas $\gamma_3$ is to be minimized ($\tau\to 0$). This contradicting criteria for selecting $\tau$ presents a design trade-off between linearity and gain. By choosing $\tau = \frac{1}{2}$ (i.e $\gamma_1 = 3$), the small signal power gain can be calculated as

\begin{align} \label{eq:G_m_small_signal_high_slope}
    G_M &= \Big[\frac{3 \pi P_I R_s}{V_{\pi} L} \Big ]^2 \frac{1}{R_s} 
\end{align}

where $s_{rmz} \triangleq \frac{3\pi P_I R_s}{V_{\pi} L}$. The optical power transfer function of this RAMZM $(\{\phi_{bias}, \theta_{DC}, \tau\} = \{\pi/2, 0, 0.5\})$ along with the transfer functions of MZMs (both single and push-pull drive) are plotted in Fig. \ref{fig:highgain}. As can be seen, the RAMZM has significantly higher slope efficiency compared to the MZMs. Here, slope efficiency is 3$\times$ (9.54 dB in power gain) better than that of MZIs driven in push-pull manner and 6$\times$ (15.56 dB in power Gain) better than that of MZMs operated in single drive \cite{cox2006analog,marpaung2009high}. This means, for the same link gain, this biasing scheme will allow upto 6$\times$ reduction in phase shifter length compared to MZMs, enabling lumped drive, smaller form factor, smaller energy footprint and excellent noise performance (discussed in Section \ref{sec:link_noise_analysis}). However, this significantly higher gain and noise performance arrives at the cost of linearity. The linearity penalty for this biasing scheme will be discussed later in Section \ref{sec:link_distortion_analysis} of this article.

\section{RAMZM Link Noise Analysis} \label{sec:link_noise_analysis}

An analog optical link using RAMZM is shown in Fig. \ref{fig:RAMZI_link}. Here, $P_{I}$ and $P_{M,o}$ are the optical power at the modulator input and output respectively. $P_{s,a}$ is the input RF power, $P_{o,d}$ and $P_{L}$ are the optical and then detected RF power at the receiver.

\begin{figure}[!t]
\centering
\includegraphics[width=\columnwidth]{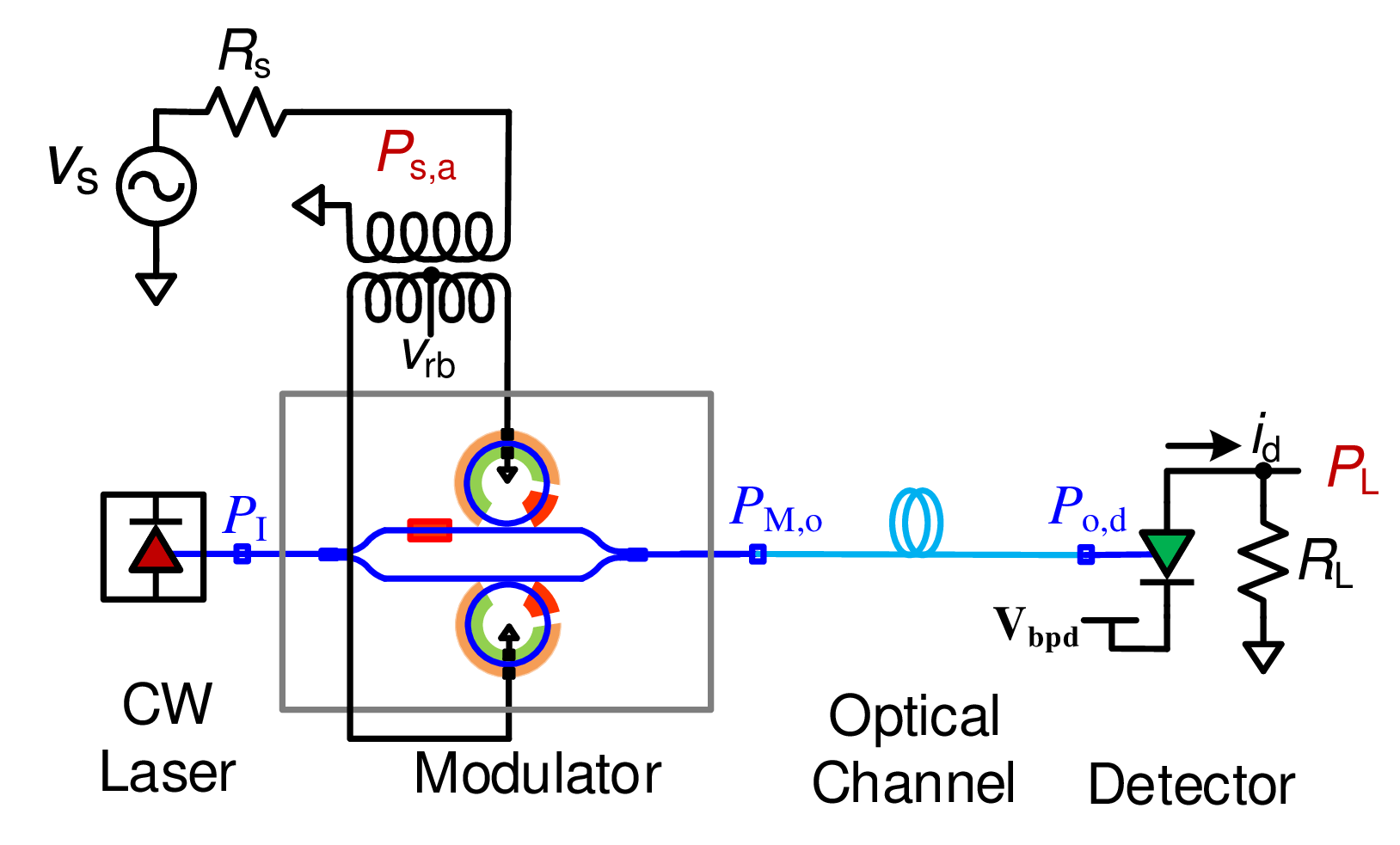}
\caption{An analog optical link using RAMZI modulator (RAMZM).}
\label{fig:RAMZI_link} 
\end{figure}

Next, the RAMZM based analog optical link gain \& noise contribution will be derived and then utilized to determine the link noise figure. Fig. \ref{fig:Link_noise_model} shows the small-signal noise model for the analog optical link seen in Fig. \ref{fig:RAMZI_link}. The link parameters used for the subsequent numerical analysis are provided in Table \ref{tab:link_param}. These link parameters are on par with the Silicon Photonic components that are available from foundry PDKs \cite{liao2018low, timurdogan2019apsuny}.

\begin{table}[hbt!]
\caption{Analog Optical Link Parameters}
\label{tab:link_param}
\begin{center}
\begin{tabular}{|l|c|c|}
\hline

Parameter & Value & Unit\\

\hline
\hline
Laser input power ($P_I$) & 13 & dBm\\
\hline
Laser Relative intensity noise (RIN) & -145 & dB/Hz\\
\hline
Modulator $V_{\pi}$ & 5 & V\\
\hline
Modulator Insertion loss (IL) & 10 & dB\\
\hline
Source Impedance ($R_s$) & 50 & Ohm\\
\hline
Detector responsivity ($r_{d}$) & 1.1 & A/W\\
\hline
Load Impedance ($R_L$) & 50 & Ohm\\
\hline
BW & 1 & Hz\\
\hline

\end{tabular}
\end{center}
\end{table}

\subsection{RAMZM Noise Modeling} 
\begin{figure}
\centering
\includegraphics[width=\columnwidth]{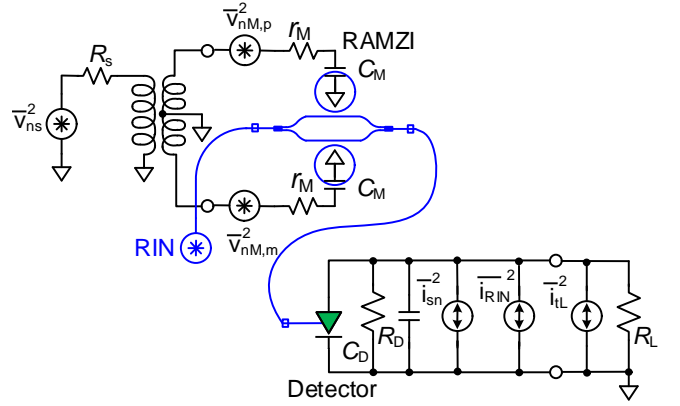}
\caption{Small-signal noise model for the analog optical link employing the RAMZM as seen in Fig. \ref{fig:RAMZI_link}.}
\label{fig:Link_noise_model}
\end{figure}

As seen in Fig. \ref{fig:Link_noise_model}, the two rings in the modulator present differential lumped capacitive load, $C_M$, with series electrode resistance, $r_M$. Thermal noise is contributed by the input source resistance ($R_s$) with noise power $kT \Delta f$. The series electrode resistances, $r_M$, also contributes thermal noise. The resulting mean square noise voltage across each $C_M$ is simply $\frac{kT}{C_M}$. The net input-referred mean square noise voltage is

\begin{equation} \label{eq:RAMZI_noise_analysis1}
    \overline{v_{n,in}^2} =  \overline{v_{nM,p}^2} + \overline{v_{nM,m}^2} 
                          =  \frac{kT}{C_M} + \frac{kT}{C_M} = \frac{2kT}{C_M} 
\end{equation}
The corresponding noise power at the RAMZM output would be $s_{rmz}^2 \frac{2kT}{R_s C_d}$. This can be interpreted as if net thermal noise is equal to the input noise power $kT \Delta f$, with the corresponding bandwidth $\Delta f = \frac{1}{2 \pi (R_s + 2r_M) \cdot C_M/2} = \frac{1}{\pi (R_s + 2r_M) \cdot C_M}$. 

\subsection{Optical Channel}
Once the RF signal is converted into the optical domain, it can be processed by optical signal processing circuits, such SiP filters \cite{choo2018automatic} and phased-arrays \cite{choo2018automatic2}, and then transported over long distances on a single-mode optical fiber. Here, we can safely assume that the optical device/channel is passive and noiseless. However, on-chip or in-package optical gain using a semiconductor optical amplifier (SOA) or erbium-doped fiber amplifier (EDFA) is also possible with the associated noise figure \cite{matsumoto2018hybrid}. 
The optical channel power gain is given as $G_{ch} \triangleq \frac{P_{o,d}}{P_{M,o}} = \frac{1}{L_{ch}}$, where $L_{ch}$ is the optical power loss.  

\subsection{Photodetector}
Photo-detection in silicon-based PICs is realized using on-chip Germanium detectors \cite{byrd2017mode}. The detector is either interfaced with a passive load $R_L$, or a transimpedance amplifier (TIA). In this work, only passive load is considered. The detected power in the load is $P_{L} = i_d^2 R_L = (r_d \cdot P_{o,d})^2 R_L$ and thus detector gain is \cite{cox2006analog} 

\begin{equation} \label{eq:G_det}
	G_{det} \triangleq \frac{P_{L}}{p_{o,d}^2} = \frac{(r_d P_{o,d})^2 R_L }{P_{o,d}^2} = r_d^2 R_L
\end{equation}
where $r_d$ is the detector responsivity. 

\subsection{RAMZM Link Gain}

In the previous section, we developed the gain of RAMZM. To be able to find the gain of the entire link, the product of modulator and detector gain must be evaluated. For simplicity, it is assumed that there is no optical loss/gain element ($\frac{p_{o,d}^2}{P_{M,o}^2}=1$) between modulator \& detector and $R_s=R_L$. Therefore, the entire link gain can be expressed as-

\begin{equation} \label{eq:G_links}
	G = \frac{P_{M,o}^2}{P_{s,a}}\cdot \frac{p_{o,d}^2}{P_{M,o}^2}\cdot\frac{P_{L}}{p_{o,d}^2} = \Big[\frac{\pi \gamma_1 P_I R_s r_d}{V_{\pi} L} \Big ]^2
\end{equation}

\begin{figure}[!h]
\centering
\includegraphics[width=\columnwidth]{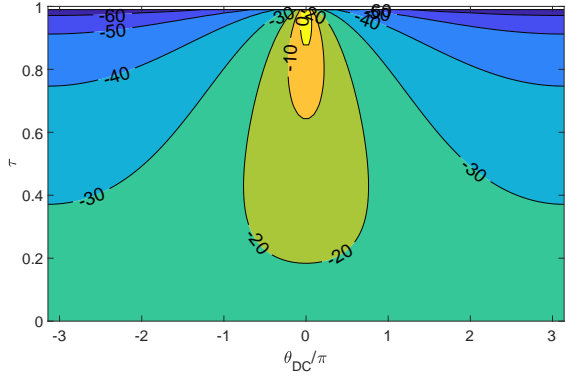}
\caption{Contour plot of RAMZM-based analog optical link Gain at different bias conditions $(\theta_{DC}, \tau)$ while keeping $\phi_{bias} = \frac{\pi}{2}$. The link parameters employed are listed in Table \ref{tab:link_param}.}
\label{fig:linkgain3d}
\end{figure}

Fig. \ref{fig:linkgain3d} shows the contour plot for the RAMZM link gain, $G$, for various bias conditions $(\theta_{DC}, \tau)$. As described in the previous section, the link gain is the highest when the thermal phase shifters on the RAMZM arm are tuned in such way that $\theta_{DC} = 0$.

\subsection{RAMZM Link Noise Figure and Ring Biasing}

The noise figure (NF) for the entire link is given by \cite{cox2006analog}

\begin{equation} \label{eq:NF2}
	NF =10 \cdot log_{10} \Big(\frac{N_{out}}{G N_i} \Big) 
\end{equation}

where $N_{out}$ is the total noise power at the link output, $N_i$ is the input noise, and $G$ is the link gain. The input noise power is due to the source resistance and given by $N_i=kT \Delta f$ \cite{cox2006analog}. By observing Fig. \ref{fig:Link_noise_model}, the noise components of the link can be summarized as \cite{cox2006analog} 

\begin{enumerate}
\item Laser relative intensity noise (RIN), $\overline{i_{rin}^2} \Delta f$ 
\item Detector shot noise, $\overline{i_{sn}^2} \Delta f$ 
\item Thermal noise from the load matched resistance of the detector ($R_L=R_{pd}$), $kT \Delta f$ 
\end{enumerate}

RIN of the laser source is expressed in dB/Hz and the resulting RIN noise current spectral density (CSD) is given by $\overline{i_{RIN}^2} = \frac{\overline{I_D}^2}{2} 10^{\frac{RIN}{10}} $ where $\overline{I_D} = r_d P_{av} = \frac{r_d P_I}{2 L}[1+cos(\phi_{bias})]$ is average detector current \cite{cox2006analog}. It is important to note that $P_{av} = \frac{P_I}{2 L}[1+cos(\phi_{bias})]$ comes from Eq. \ref{eq:Pmo1}, weighted by optical loss L. The detector shot noise current spectral density is described as $\overline{i_{sn}^2} = 2q \overline{I_D}$. Thus, RIN and shot noise CSD depend upon $\overline{I_D^2}$ and $\overline{I_D}$ respectively. 

When lossy impedance matching is implemented at the input and output side, the total noise for the link is derived as


\begin{equation} \label{eq:NF_link1}
	N_{out}= \frac{1}{4}(\overline{i_{rin}^2} + \overline{i_{sn}^2})\Delta f R_L + (1+G)kT \Delta f 		
\end{equation}

Consequently, the NF of the link is expressed as

\begin{equation} \label{eq:NF_primary}
	NF = 10\cdot log_{10} \Big[1 + \frac{R_L}{4kTG} \Big(\frac{\overline{I_D}^2}{2} 10^{\frac{RIN}{10}} + 2q \overline{I_D} \Big) + \frac{1}{G} \Big]	
\end{equation}

\begin{figure}[!h]
\centering
\includegraphics[width=\columnwidth]{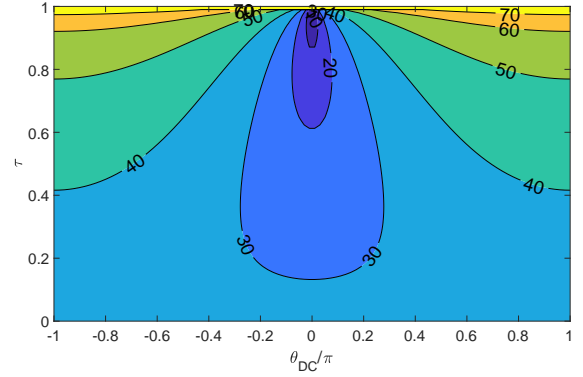}
\caption{Contour plot of RAMZM-based analog optical link NF at different bias conditions $(\theta_{DC}, \tau)$ while keeping $\phi_{bias} = \frac{\pi}{2}$. The link parameters employed are the same as in Table \ref{tab:link_param}.}
\label{fig:nfplot3d}
\end{figure}

The corresponding plot of NF for the different bias conditions $(\theta_{DC}, \tau)$ is illustrated in Fig. \ref{fig:nfplot3d}. Here, it is evident that for $\theta_{DC} = 0$, NF is maximized due to the large link gain.

On the other hand, if a passive transformer with turn ratio, $N_D$, is used to match diode to the load, $R_L$,  link NF in Eq. \ref{eq:NF2} is derived as-
\begin{equation} \label{eq:NF_link2}
	NF = 10 \cdot log_{10} \Big[1 + \frac{N_D^2R_s}{kTG} \Big(\frac{\overline{I_D}^2}{2} 10^{\frac{RIN}{10}} + 2q \overline{I_D} \Big) + \frac{1}{G} \Big]
\end{equation}



\begin{figure}[!h]
\centering
\includegraphics[width=\columnwidth]{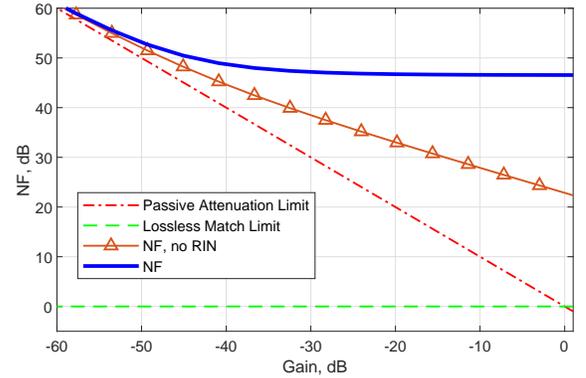}
\caption{NF versus gain for a RAMZM-based analog optical link. Here, $P_I$ and $L_{ch}$ are varied to sweep the gain. Also, $P_I<30dBm$ while other link parameters are from Table \ref{tab:link_param}.}
\label{fig:link_analysis1}
\end{figure}

Fig. \ref{fig:link_analysis1} plots Eq. \ref{eq:NF_link2} with respect to the small-signal power gain, $G$. If the detector noise terms are ignored, then for large optical attenuation (i.e. $G \ll 0~dB$), the load thermal noise dominates the detector noise terms and Eq. \ref{eq:NF_link2} reduces to the \textit{passive attenuation limit} $NF = log_{10} \Big(1+\frac{1}{G} \Big)$. For $G \gg 0~dB$, the NF asymptotically approaches the \textit{lossless match limit} of 0~dB. One can observe that the link NF is rather high due to the detector noise, as $\overline{I_D}$ depends upon $\frac{P_I}{2 L}[1+cos(\phi_{bias})]$. This can be improved by biasing the arms of the modulator at different angle, $\phi_{bias}> \frac{\pi}{2}$ \cite{cox2006analog,marpaung2009high}, or by filtering out the optical carrier \cite{lagasse1994optical,marpaung2009high} using an on-chip SiP filter \cite{shawon2020rapid}.

As discussed before, the slope efficiency of RAMZM, when perfect third-order nonlinearity cancellation is achieved, is $\frac{1}{3}$ (or 9.54 dB lower power gain) compared to the MZMs (push-pull drive). This lower gain also result in higher NF in RAMZM, approximately by the same proportion. On the other hand, the gain enhanced RAMZM will have 9.54 dB improvements in both power gain and NF.

\subsection{Null Biasing of RAMZM}

\begin{figure}[!h]
\centering
\includegraphics[width=1\columnwidth]{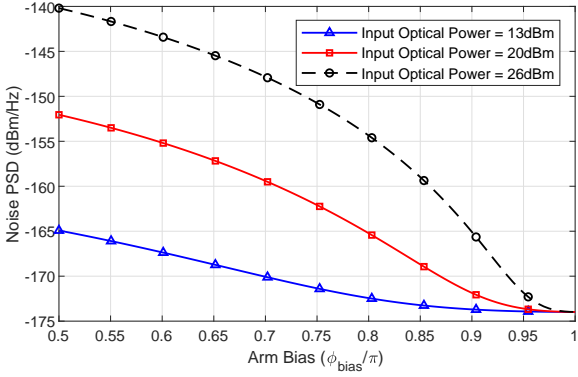}
\caption{RAMZM Noise PSD vs arm bias angle ($\phi_{bias}$) for different input optical power, $P_I$. Other link parameters employed are from Table \ref{tab:link_param}.}
\label{fig:noisepsd}
\end{figure}

As discussed earlier, RIN and shot noise CSD depend upon $\overline{I_D^2}$ and $\overline{I_D}$ respectively. This means, reducing $\overline{I_D}$ improves noise performance of the link, especially when the link is RIN limited ($\overline{I_D^2}$ dependence). Since $\overline{I_D} = \frac{r_d P_I}{2 L}[1+cos(\phi_{bias})]$, by changing $\phi_{bias}$ (null or low biasing \cite{farwell1993increased, sisto2006gain}) we can reduce the overall noise as shown in Fig. \ref{fig:noisepsd}.

\begin{figure}[!h]
\centering
\includegraphics[width=1\columnwidth]{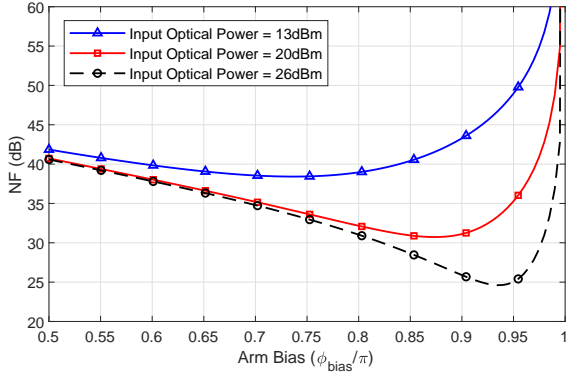}
\caption{RAMZM NF vs arm bias angle ($\phi_{bias}$) for different input optical power. The link parameters for these plots are from Table \ref{tab:link_param}.}
\label{fig:lowbias3}
\end{figure}

Although the link gain reduces along with noise in null biasing technique (also known as low-biasing in the context of MZMs), the overall advantage in NF can be achieved, especially when a high laser power ($P_I$) is used as illustrated in Fig. \ref{fig:lowbias3}.

\begin{figure}[!h]
\centering
\includegraphics[width=1\columnwidth]{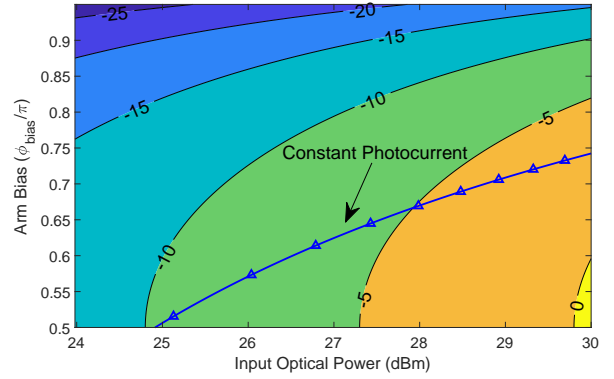}
\caption{Contour plot of Link Gain for various arm bias ($\phi_{bias}$) and input optical power. The link parameters for the analysis are from Table \ref{tab:link_param}.}
\label{fig:lowbiasing}
\end{figure}

On the other hand, although mode evolution based photodiodes (PDs) provide higher saturation power over butt-coupled PDs \cite{byrd2017mode}, the use of high optical input power in integrated photonics is still limited due to poor saturation power performances of the PDs. In such cases, null biasing can be implemented in the RAMZM based analog optical link. When the link gain is limited by PD saturation, by employing null biasing one can increase the link gain without increasing PD current just by increasing the laser power (as shown in Fig. \ref{fig:lowbiasing}).

In Fig. \ref{fig:lowbiasing}, a constant photocurrent of 15.5mA \cite{byrd2017mode} is plotted over the input laser power and RAMZM arm bias versus Gain. It can be observed that by increasing laser power from 25 to 30dBm, it's possible to achieve a Link Gain advantage of $>$6dB when RAMZM arm bias is adjusted without having to increase the PD current.

\section{RAMZM Link Distortion Analysis} \label{sec:link_distortion_analysis}

Fig. \ref{fig:RAMZI_link} had two active components that contribute distortion in the RF photonic link, namely the modulator and the detector. Unlike RAMZM, detector distortion does not arise from the optical to photo-current transfer characteristics, but from the higher-order effects \cite{cox2006analog}. These experimentally characterized effects depend upon the reverse bias voltage, $V_{b,pd}$, and  $\overline{I_D}$, and can be minimized by using a sufficiently large $V_{b,pd}$ and avoiding detector saturation (by restricting $\overline{I_D}$ using null biasing technique) \cite{cox2006analog}. Since the distortion in an analog photonic link is dominated by the modulator nonlinearity, this study of link distortion analysis only focuses on RAMZM. Also, since harmonic distortions only appear in very broadband links, here we limit our discussion only to intermodulation distortions.

\subsection{Second and Third-order Intercept Point}
From Eq. \ref{eq:Pmo1}, the optical power of the third-order intermodulation distortion for a two-tone input, $\theta_{mod} = \frac{\pi V_m}{V_{\pi}} \cdot [sin(\omega_1 t) + sin(\omega_2 t)]$, can be written as  $P_{optical-im3} = \frac{P_I}{L} \cdot \frac{3\gamma_3}{4} \cdot (\frac{\pi V_m}{V_{\pi}})^3 cos(2\omega_1 \pm \omega_2)$ at $2\omega_1 \pm \omega_2$ \cite{cox2006analog}. On the other hand, its fundamental counterpart can be expressed as $P_{optical-fund} = \frac{P_I}{L} \cdot \gamma_1 \cdot (\frac{\pi V_m}{V_{\pi}}) cos(\omega_1)$ at $\omega_1$. Since a lossy impedance matching is considered at the photodiode output, the output electrical powers are expressed as

\begin{equation} \label{eq:pfund}
P_{fund} = \big(\frac{1}{2} r_d P_{optical-fund}\big)^2 R_L
\end{equation}

\begin{equation} \label{eq:pim3}
P_{im3} = \big(\frac{1}{2} r_d P_{optical-im3}\big)^2 R_L
\end{equation}

Using $V_m = \sqrt{2P_{in}R_s}$ in Eq. \ref{eq:pfund} and \ref{eq:pim3}, we can express all electrical output powers in terms of input electrical power, $P_{in}$ as-

\begin{equation} \label{eq:pfundpin}
P_{fund} = \Big(\frac{\pi^2 r_d^2 P_I^2 R_s R_L\gamma_1^2}{4V_{\pi}^2 L^2}\Big)\cdot P_{in}
\end{equation}

\begin{equation} \label{eq:pim3pin}
P_{im3} = \Big(\frac{9\pi^6 r_d^2 P_I^2 R_s^3 R_L\gamma_3^2}{16V_{\pi}^6 L^2}\Big)\cdot P_{in}^3
\end{equation}

Now, the $P_{in}$ for which $P_{fund} = P_{im3}$ is called the third order intercept point (IIP3) \cite{kundert2002accurate}. Therefore, equating Eqs. \ref{eq:pfundpin} and \ref{eq:pim3pin} and solving for $P_{in}$ results in

\begin{equation} \label{eq:iip3}
IIP_3 = \frac{2V_{\pi}^2\big[\tau^2 - 2\cos(\theta_{DC})\tau + 1\big]^2}{\pi^2 R_s \big[2\cos(\theta_{DC})^2 \tau^2 + (\tau^3 + \tau)\cos(\theta_{DC}) + 2\tau^4 - 8\tau^2 + 2\big]}
\end{equation}

Similarly, it can be shown that

\begin{equation} \label{eq:iip2}
IIP_2 = \frac{V_{\pi}^2\big[\tau^2 - 2\cos(\theta_{DC})\tau + 1\big]^2}{2 \pi^2 R_s (\tau^2-1)^2}\cdot \tan(\phi_{bias})^2
\end{equation}

From these expressions, it can be noted that both IIP$_2$ and IIP$_3$ are strong functions of $V_{\pi}$. Also, the IIP$_3$ of an RAMZM is independent of the arm bias ($\phi_{bias}$), but strongly dependent upon ring bias  ($\theta_{DC}$) and the coupling ratio. On the other hand, unlike IIP$_3$, IIP$_2$ strongly depends upon arm biasing.

\begin{figure}[!h]
\centering
\includegraphics[width=1\columnwidth]{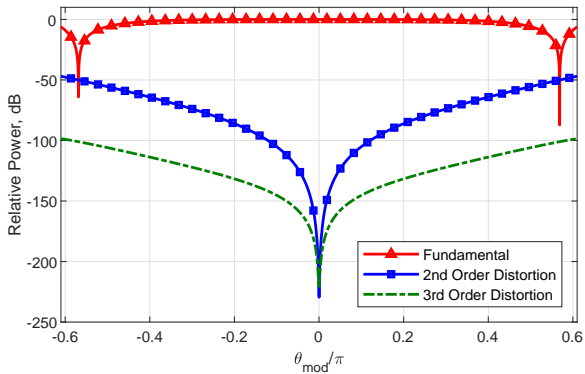}
\caption{RAMZM Fundamental, second and third-order distortion components \textit{vs} ring modulation angle, $\theta_{mod}$. Here, the RAMZM is biased at $\{\phi_{bias}, \theta_{DC}, \tau \}=\{\frac{\pi}{2},\pi, \frac{1}{2}\}$}
\label{fig:imd3_bias}
\end{figure}

To show the dynamic characteristics, relative power of the fundamental, IM$_2$ and IM$_3$ are plotted against the modulation angle of the RAMZM in Fig. \ref{fig:imd3_bias}. Here, 1st, 2nd and 3rd derivatives of the RAMZM transfer function are calculated to find the relative power of the fundamental, IM$_2$ and IM$_3$ distortions considering $\{\phi_{bias}, \theta_{DC}, \tau \}=\{\frac{\pi}{2},\pi, \frac{1}{2}\}$. Here, we see that for small modulation stimulus, the IM$_2$ and IM$_3$ are virtually non-existent.

\begin{figure}[!h]
\centering
\includegraphics[width=\columnwidth]{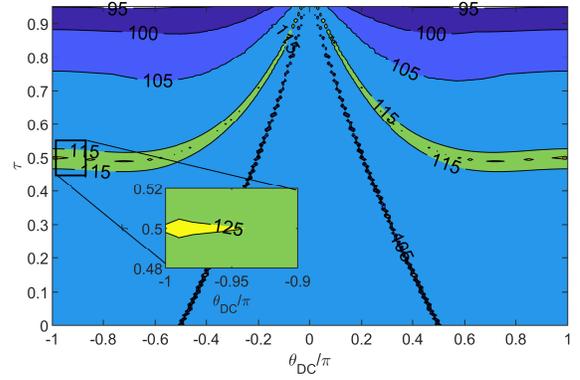}
\caption{Contour plot of RAMZM-based analog optical link SFDR in $dB/Hz^{\frac{2}{3}}$ at different bias conditions, keeping $\phi_{bias} = \frac{\pi}{2}$. The link parameters employed are listed in Table \ref{tab:link_param}.}
\label{fig:sfdr3dfinal}
\end{figure}

To provide a comprehensive view of the link gain and total noise, SFDR is evaluated and plotted in Fig. \ref{fig:sfdr3dfinal}) using the method described in \cite{kundert2002accurate} and link parameters listed in Table \ref{tab:link_param}. Here, we see that the highest SFDR of $\sim 128$ dB/Hz$^{2/3}$ is obtained when the rings are biased in anti-resonance and $\tau=\frac{1}{2}$. As ring bias are tuned away from anti-resonance, a narrow band for coupling ratio $\tau$ exists for highest SFDR. The generated countours in Figs. \ref{fig:linkgain3d}, \ref{fig:nfplot3d} and \ref{fig:sfdr3dfinal} can be used together to make design trade-offs between the link NF and distortion.

\subsection{Linearity of RAMZM and MZM}

\begin{figure}[!h]
\centering
\includegraphics[width=1\columnwidth]{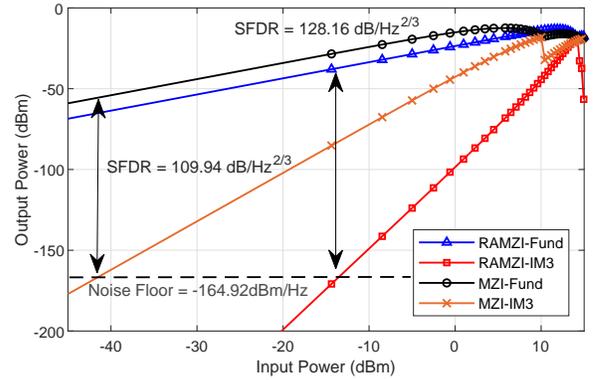}
\caption{SFDR performance of RAMZM and MZM in a Photonic Link. Here, RAMZM and MZM bias conditions are $\{\phi_{bias}, \theta_{DC}, \tau \}=\{\frac{\pi}{2},\pi, \frac{1}{2}\}$ and $\{\phi_{bias}\}=\{\frac{\pi}{2}\}$, respectively. Also, the link parameters are from Table \ref{tab:link_param}.}
\label{fig:ramzmvsmzmsfdr}
\end{figure}

As discussed before, MZM suffers from higher nonlinearity due to its sinusoidal transfer function. To quantify the linearity of both MZM and linearized RAMZM, $SFDR_3$ is calculated in Fig. \ref{fig:ramzmvsmzmsfdr}. Here, RAMZM and MZM have SFDR values of 128.16 and 109.94 $dB/Hz^{\frac{2}{3}}$ at 1GHz, respectively. This is 18dB improvement over regular MZMs, which makes RAMZM very attractive for high performance RF photonic applications.

\subsection{Linearity of Gain Enhanced (GE) RAMZM}

\begin{figure}[!h]
\centering
\includegraphics[width=1\columnwidth]{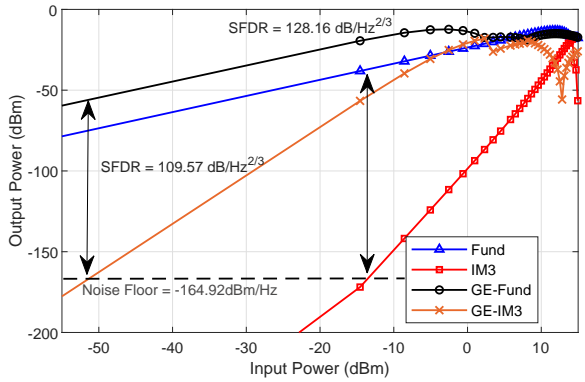}
\caption{Comparison of SFDR performance of RAMZMs in a Photonic Link with different bias conditions. Here, Linearized and Gain Enhanced (GE) RAMZM bias conditions are $\{\phi_{bias}, \theta_{DC}, \tau \}=\{\frac{\pi}{2},\pi, \frac{1}{2}\}$ and $\{\frac{\pi}{2},0, \frac{1}{2}\}$, respectively. Also, the link parameters are from Table \ref{tab:link_param}.}
\label{fig:ramzivsramzi}
\end{figure}

In section II, we proposed a new biasing scheme where slope efficiency of RAMZM is upto 6$\times$ higher than that of MZMs. However, as previously discussed, this gain enhancement comes at the expense of linearity since $\theta_{DC}$ has to be shifted from $\pi$ towards 0. In Fig. \ref{fig:ramzivsramzi}, we show that the SFDR of GE RAMZM is about 109.57 $dB/Hz^{\frac{2}{3}}$ at 1GHz. It is interesting to note that the GE RAMZM provides upto 6$\times$ improvement in slope efficiency over MZM while still providing the same SFDR performance as an MZM.

\section{Conclusion}
In this work, we presented analysis for silicon photonic RAMZM-based analog optical link that provides significant improvement in linearity over the MZM-based links. It is shown that the linearized RAMZM-based link provides 18dB/H$z^{\frac{2}{3}}$ higher SFDR than its MZM counterpart. Using analytical expressions, we have also shown how link and RAMZM parameters affect the link Gain, Noise, Linearity performances. The associated link design trade-offs are also explored. Using an on-chip RAMZM, fabricated in IMEC’s ISIPP50G Silicon Photonic process, the multi-biasing process is described and explained. We have also proposed a novel biasing scheme which improves link gain and NF by upto 15.5dB (compared to MZM-based links). This will enable the use of smaller depletion mode/forward biased phase-shifters (lumped capacitive element) in Silicon Photonic technology. This will significantly reduce the complexity of modulator driver design since transmission line drive will not be required. Moreover, null biasing technique is explored for photodiode saturation limited RAMZM-based optical link. This study will provide the designers with in-depth understanding into the design tradeoffs and nuances of high performance Silicon Photonic modulator based link design.

\section{Acknowledgment}
The authors gratefully acknowledge the generous funding support from the Air Force Office of Sponsored Research (AFOSR) YIP Award FA9550-
17-1-0076 and the National Science Foundation (NSF) CAREER Award EECS-2014109.



%


\ifCLASSOPTIONcaptionsoff
  \newpage
\fi


\bibliographystyle{IEEEtran}
\bibliography{RAMZM_LINK}

\end{document}